\documentclass[11pt]{article}
\usepackage{footnote}
\usepackage[dvips]{graphicx}
\usepackage{epsfig}
\usepackage{amssymb}
\usepackage{color}
\usepackage[left]{lineno}

\begin{document}
\centerline{\LARGE EUROPEAN ORGANIZATION FOR NUCLEAR RESEARCH}
%
%
\vspace{10mm} {\flushright{
CERN-PH-EP-2014-025 \\
17 February 2014\\
Revised version:\\11 March 2014\\
}}
\vspace{-30mm}

%
\vspace{40mm}

\begin{center}
\boldmath
{\bf {\Large \boldmath{Study of the $K^\pm\to\pi^\pm\gamma\gamma$ decay by the NA62 experiment}}}
\unboldmath
\end{center}
\begin{center}
{\Large The NA62 collaboration}\\
\end{center}

\begin{abstract}
A study of the dynamics of the rare decay $K^\pm\to\pi^\pm\gamma\gamma$ has been performed on a sample of 232 decay candidates, with an estimated background of $17.4\pm1.1$ events, collected by the NA62 experiment at CERN in 2007. The results are combined with those from a measurement conducted by the NA48/2 collaboration at CERN. The combined model-independent branching ratio in the kinematic range $z=(m_{\gamma\gamma}/m_K)^2>0.2$ is ${\cal B}_{\rm MI}(z>0.2) = (0.965 \pm 0.063) \times 10^{-6}$, and the combined branching ratio in the full kinematic range assuming a Chiral Perturbation Theory description is ${\cal B}(K_{\pi\gamma\gamma}) = (1.003 \pm 0.056) \times 10^{-6}$. A detailed comparison of the results with the previous measurements is performed.
\end{abstract}

\begin{center}
{\it Accepted for publication in Physics Letters B}
\end{center}

\newpage
\begin{center}
{\Large The NA62 collaboration}\\
\vspace{2mm}
 C.~Lazzeroni$\,$\footnotemark[1],
 A.~Romano\\
{\em \small University of Birmingham, Edgbaston, Birmingham, B15 2TT, United Kingdom} \\[0.2cm]
 A.~Ceccucci,
 H.~Danielsson,
 V.~Falaleev,
 L.~Gatignon,
 S.~Goy Lopez$\,$\footnotemark[2], \\
 B.~Hallgren$\,$\footnotemark[3],
 A.~Maier,
 A.~Peters,
 M.~Piccini$\,$\footnotemark[4],
 P.~Riedler\\
{\em \small CERN, CH-1211 Gen\`eve 23, Switzerland} \\[0.2cm]
 P.L.~Frabetti,
 E.~Gersabeck$\,$\footnotemark[5],
 V.~Kekelidze,
 D.~Madigozhin,
 M.~Misheva, \\
 N.~Molokanova,
 S.~Movchan,
 Yu.~Potrebenikov,
 S.~Shkarovskiy,
 A.~Zinchenko\\
{\em \small Joint Institute for Nuclear Research, 141980 Dubna (MO), Russia} \\[0.2cm]
 P.~Rubin$\,$\footnotemark[6]\\
{\em \small George Mason University, Fairfax, VA 22030, USA} \\[0.2cm]
 W.~Baldini,
 A.~Cotta Ramusino,
 P.~Dalpiaz,
 M.~Fiorini,
 A.~Gianoli, \\
 A.~Norton,
 F.~Petrucci,
 M.~Savri\'e,
 H.~Wahl\\
{\em \small Dipartimento di Fisica e Scienze della Terra dell'Universit\`a e Sezione
dell'INFN di Ferrara, \\ I-44122 Ferrara, Italy} \\[0.2cm]
 A.~Bizzeti$\,$\footnotemark[7],
 F.~Bucci$\,$\footnotemark[8],
 E.~Iacopini$\,$\footnotemark[8],
 M.~Lenti,
 M.~Veltri$\,$\footnotemark[9]\\
{\em \small Sezione dell'INFN di Firenze, I-50019 Sesto Fiorentino, Italy} \\[0.2cm]
 A.~Antonelli,
 M.~Moulson,
 M.~Raggi,
 T.~Spadaro \\
{\em \small Laboratori Nazionali di Frascati, I-00044 Frascati, Italy}\\[0.2cm]
 K.~Eppard,
 M.~Hita-Hochgesand,
 K.~Kleinknecht,
 B.~Renk,
 R.~Wanke,
 A.~Winhart$\,$\footnotemark[3]\\
{\em \small Institut f\"ur Physik, Universit\"at Mainz, D-55099 Mainz, Germany$\,$\footnotemark[10]} \\[0.2cm]
 R.~Winston\\
{\em \small University of California, Merced, CA 95344, USA} \\[0.2cm]
 V.~Bolotov$\,$\renewcommand{\thefootnote}{\fnsymbol{footnote}}%
\footnotemark[2]\renewcommand{\thefootnote}{\arabic{footnote}},
 V.~Duk$\,$\footnotemark[4],
 E.~Gushchin\\
{\em \small Institute for Nuclear Research, 117312 Moscow, Russia} \\[0.2cm]
 F.~Ambrosino,
 D.~Di Filippo,
 P.~Massarotti,
 M.~Napolitano,
 V.~Palladino$\,$\footnotemark[11],
 G.~Saracino \\
{\em \small Dipartimento di Fisica dell'Universit\`a e Sezione dell'INFN di Napoli, I-80126 Napoli, Italy}\\[0.2cm]
 G.~Anzivino,
 E.~Imbergamo,
 R.~Piandani$\,$\footnotemark[12],
 A.~Sergi$\,$\footnotemark[3]\\
{\em \small Dipartimento di Fisica dell'Universit\`a e Sezione dell'INFN di Perugia, I-06100 Perugia, Italy} \\[0.2cm]
 P.~Cenci,
 M.~Pepe\\
{\em \small Sezione dell'INFN di Perugia, I-06100 Perugia, Italy} \\[0.2cm]
 F.~Costantini,
 N.~Doble,
 S.~Giudici,
 G.~Pierazzini$\,$\renewcommand{\thefootnote}{\fnsymbol{footnote}}%
\footnotemark[2]\renewcommand{\thefootnote}{\arabic{footnote}},
 M.~Sozzi,
 S.~Venditti$\,$\footnotemark[11]\\
{\em Dipartimento di Fisica dell'Universit\`a e Sezione dell'INFN di Pisa, I-56100 Pisa, Italy} \\[0.2cm]
 S.~Balev$\,$\renewcommand{\thefootnote}{\fnsymbol{footnote}}%
\footnotemark[2]\renewcommand{\thefootnote}{\arabic{footnote}},
 G.~Collazuol$\,$\footnotemark[13],
 L.~DiLella,
 S.~Gallorini$\,$\footnotemark[13],
 E.~Goudzovski$\,$\renewcommand{\thefootnote}{\fnsymbol{footnote}}%
\footnotemark[1]\renewcommand{\thefootnote}{\arabic{footnote}}$^,$\footnotemark[1]$^,$\footnotemark[3], \\
 G.~Lamanna,
 I.~Mannelli,
 G.~Ruggiero$\,$\footnotemark[11]\\
{\em Scuola Normale Superiore e Sezione dell'INFN di Pisa, I-56100 Pisa, Italy} \\[0.2cm]
 C.~Cerri,
 R.~Fantechi \\
{\em Sezione dell'INFN di Pisa, I-56100 Pisa, Italy} \\[0.2cm]
 S.~Kholodenko,
 V.~Kurshetsov,
 V.~Obraztsov,
 V.~Semenov,
 O.~Yushchenko\\
{\em \small Institute for High Energy Physics, 142281 Protvino (MO), Russia}$\,$\footnotemark[14] \\[0.2cm]
\newpage
 G.~D'Agostini\\
{\em \small Dipartimento di Fisica, Sapienza Universit\`a di Roma and\\Sezione dell'INFN di Roma I, I-00185 Roma, Italy} \\[0.2cm]
 E.~Leonardi,
 M.~Serra,
 P.~Valente\\
{\em \small Sezione dell'INFN di Roma I, I-00185 Roma, Italy} \\[0.2cm]
 A.~Fucci,
 A.~Salamon\\
{\em \small Sezione dell'INFN di Roma Tor Vergata, I-00133 Roma, Italy} \\[0.2cm]
 B.~Bloch-Devaux$\,$\footnotemark[15],
 B.~Peyaud\\
{\em \small DSM/IRFU -- CEA Saclay, F-91191 Gif-sur-Yvette, France} \\[0.2cm]
 J.~Engelfried\\
{\em \small Instituto de F\'isica, Universidad Aut\'onoma de San
Luis Potos\'i, 78240 San Luis Potos\'i, Mexico}$\,$\footnotemark[16] \\[0.2cm]
 D.~Coward\\
{\em \small SLAC National Accelerator Laboratory, Stanford University, Menlo Park, CA 94025, USA} \\[0.2cm]
 V.~Kozhuharov$\,$\footnotemark[17],
 L.~Litov \\
{\em \small Faculty of Physics, University of Sofia, 1164 Sofia, Bulgaria}$\,$\footnotemark[18] \\[0.2cm]
 R.~Arcidiacono$\,$\footnotemark[19],
 S.~Bifani$\,$\footnotemark[3] \\
{\em \small Dipartimento di Fisica dell'Universit\`a e
Sezione dell'INFN di Torino, I-10125 Torino, Italy} \\[0.2cm]
 C.~Biino,
 G.~Dellacasa,
 F.~Marchetto \\
{\em \small Sezione dell'INFN di Torino, I-10125 Torino, Italy} \\[0.2cm]
 T.~Numao,
 F.~Reti\`{e}re \\
{\em \small TRIUMF, 4004 Wesbrook Mall, Vancouver, British Columbia, V6T 2A3, Canada} \\[0.2cm]
\end{center}
%
%
\renewcommand{\thefootnote}{\fnsymbol{footnote}}
\footnotetext[1]{Corresponding author, email: eg@hep.ph.bham.ac.uk}
\footnotetext[2]{Deceased}
\renewcommand{\thefootnote}{\arabic{footnote}}
\footnotetext[1]{Supported by a Royal Society University Research Fellowship}
\footnotetext[2]{Present address: CIEMAT, E-28040 Madrid, Spain}
\footnotetext[3]{Present address: School of Physics and Astronomy, University of Birmingham, Birmingham, B15 2TT, UK}
\footnotetext[4]{Present address: Sezione dell'INFN di Perugia, I-06100 Perugia, Italy}
\footnotetext[5]{Present address: Ruprecht-Karls-Universit\"{a}t Heidelberg, D-69120 Heidelberg, Germany}
\footnotetext[6]{Funded by the National Science Foundation under award No. 0338597}
\footnotetext[7]{Also at Dipartimento di Fisica, Universit\`a di Modena e Reggio Emilia, I-41125 Modena, Italy}
\footnotetext[8]{Also at Dipartimento di Fisica, Universit\`a di Firenze, I-50019 Sesto Fiorentino, Italy}
\footnotetext[9]{Also at Istituto di Fisica, Universit\`a di Urbino, I-61029 Urbino, Italy}
\footnotetext[10]{Funded by the German Federal Minister for Education and Research (BMBF) under contract 05HA6UMA}
\footnotetext[11]{Present address: CERN, CH-1211 Gen\`eve 23, Switzerland}
\footnotetext[12]{Present address: Sezione dell'INFN di Pisa, I-56100 Pisa, Italy}
\footnotetext[13]{Present address: Dipartimento di Fisica dell'Universit\`a e Sezione dell'INFN di Padova, I-35131 Padova, Italy}
\footnotetext[14]{Partly funded by the Russian Foundation for Basic Research grant 12-02-91513}
\footnotetext[15]{Present address: Dipartimento di Fisica dell'Universit\`a di Torino, I-10125 Torino, Italy}
\footnotetext[16]{Funded by Consejo Nacional de Ciencia y Tecnolog\'{\i}a {\nobreak (CONACyT)} and Fondo de Apoyo a la Investigaci\'on (UASLP)}
\footnotetext[17]{Present address: Laboratori Nazionali di Frascati, I-00044 Frascati, Italy}
\footnotetext[18]{Funded by the Bulgarian National Science Fund under contract DID02-22}
\footnotetext[19]{Present address: Universit\`a degli Studi del Piemonte Orientale, I-13100 Vercelli, Italy}

\newpage


\section*{Introduction}

Experimental studies of radiative non-leptonic kaon decays allow crucial tests of Chiral Perturbation Theory (ChPT) describing weak low energy processes: the first non-trivial contribution to their decay rates stems from next-to-leading order ChPT. For the rare decay $K^\pm\to\pi^\pm\gamma\gamma$ ($K_{\pi\gamma\gamma}$), considerable phenomenological understanding~\cite{se72, ec88, da96, ge05} is not matched by sufficient experimental data.

The $K_{\pi\gamma\gamma}$ decay can be described by two kinematic variables:
\begin{displaymath}
z = \frac{(q_1+q_2)^2}{m_K^2} = \left(\frac{m_{\gamma\gamma}}{m_K}\right)^2, ~~~
y = \frac{p(q_1-q_2)}{m_K^2}~,
\end{displaymath}
where $p$ and $q_{1,2}$ are the 4-momenta of the kaon and the two photons respectively, $m_{\gamma\gamma}$ is the di-photon invariant mass, and $m_K$ is the charged kaon mass. The allowed region of the kinematic variables is~\cite{da96}
\begin{displaymath}
0\le z \le z_{\rm max}=(1-r_\pi)^2=0.515, ~~~~ 0\le y\le y_{\rm max}(z) = \frac{1}{2}\sqrt{\lambda\left(1, r_\pi^2,z\right)},
\end{displaymath}
where $r_\pi=m_\pi/m_K$, $m_\pi$ is the charged pion mass and $\lambda(a,b,c)=a^2+b^2+c^2-2(ab+ac+bc)$.

The $K_{\pi\gamma\gamma}$ decay was first observed by the BNL E787 experiment in 1997~\cite{ki97}: 31 $K^+$ decay candidates were reported in the kinematic region $100~{\rm MeV}/c<p_\pi^*<180~{\rm MeV}/c$ or $0.157<z<0.384$ ($p_\pi^*$ is the $\pi^+$ momentum in the $K^+$ rest frame). An analysis of 149 $K^\pm_{\pi\gamma\gamma}$ decay candidates in the kinematic region $z>0.2$ was published by the NA48/2 collaboration at CERN in 2014~\cite{ba14}. A related decay mode $K^\pm\to\pi^\pm\gamma e^+e^-$ ($K_{\pi\gamma ee}$) has been measured from 120 candidates in the kinematic region $m_{\gamma ee}>260~{\rm MeV}/c^2$ or $z=(m_{\gamma ee}/m_K)^2>0.277$ at the NA48/2 experiment~\cite{ba08}.

A measurement of the $K_{\pi\gamma\gamma}^\pm$ decay based on a minimum bias data sample collected by the NA62 experiment in 2007 is reported here. The results are combined with those from the NA48/2 measurement~\cite{ba14} and compared to the earlier measurements~\cite{ki97,ba08}.

\section{Beam and detector}
\label{sec:beam}

The beam line and setup of the earlier NA48/2 experiment~\cite{ba07} were used for the NA62 data taking in 2007. However, the beam line parameters and transverse momentum kick provided by the spectrometer magnet were significantly different. Secondary beams of positive and negative hadrons with a central momentum of 74 GeV/$c$ and a momentum spread of $1.4$ GeV/$c$ (rms) were derived from the primary 400 GeV/$c$ protons extracted from the CERN SPS and interacting with a beryllium target. These beams were dominated by $\pi^\pm$; the $K^\pm$ component was about 6\%. They were delivered, either alternately or simultaneously, into a 114~m long cylindrical vacuum tank containing the fiducial decay region at angles of $\pm0.23$~mrad with respect to the detector axis, so as to compensate for the opposite $\mp3.58$~mrad deflections by the downstream spectrometer magnet. These deflections were regularly reversed during data taking. The fraction of beam kaons decaying in the vacuum tank was $18\%$.

The momenta of charged decay products of $K^\pm$ were measured by a magnetic spectrometer, housed in a tank filled with helium at nearly atmospheric pressure, located downstream of the decay vacuum tank and separated from it by a thin ($0.3\%X_0$) $\rm{Kevlar}\textsuperscript{\textregistered}$ composite window. An aluminium beam pipe of 158~mm outer diameter and 1.1~mm thickness traversing the centre of the spectrometer (and all the following detectors) allowed the undecayed beam particles and the muon halo from beam pion decays to continue their path in vacuum. The spectrometer consisted of four drift chambers (DCHs) with a radial extension of 1.35~m, and a dipole magnet located between the second and the third DCH, which provided a horizontal momentum kick of $265~\mathrm{MeV}/c$. The nominal spectrometer momentum resolution was $\sigma_p/p = (0.48\oplus 0.009\cdot p)\%$, where the momentum $p$ is expressed in GeV/$c$. A hodoscope (HOD) consisting of two planes of 64 plastic scintillator strips, with each plane arranged in four quadrants, was placed downstream of the spectrometer. The HOD provided trigger signals and time measurements of charged particles with a resolution of about 150~ps. Following the hodoscope was a quasi-homogeneous liquid krypton electromagnetic calorimeter (LKr) with an active volume of 7 m$^3$, $27X_0$ deep, segmented transversally into 13248 projective $\sim\!2\!\times\!2$~cm$^2$ cells (with no longitudinal segmentation). The LKr energy resolution was $\sigma_E/E=(3.2/\sqrt{E}\oplus9/E\oplus0.42)\%$, and its spatial resolution for the transverse coordinates $x$, $y$ of an isolated electromagnetic shower was $\sigma_x=\sigma_y=(4.2/\sqrt{E}\oplus0.6)$~mm, where $E$ is expressed in GeV. A plane of scintillating fibres located in the LKr calorimeter volume at a depth of about $9.5X_0$, close to the maxima of showers initiated by 10~GeV photons, formed the ``neutral hodoscope'' (NHOD) which also provided trigger signals. The LKr was followed by a hadronic calorimeter and a muon detector, neither being used in the present analysis. A detailed description of the detector can be found in Ref.~\cite{fa07}.

\section{Data sample and trigger}
\label{sec:sample}

The data were obtained from about $3.5\times 10^5$ SPS spills recorded during 4 months of operation in 2007 with low intensity beams at an instantaneous kaon decay rate in the vacuum tank of $\sim10^5$~Hz. The total number of $K^\pm$ decays in the vacuum tank was $\sim 2 \times 10^{10}$. About 27\% of them were collected with simultaneous $K^+$ and $K^-$ beams, with a $K^+$/$K^-$ flux ratio of 2.0 and
an angle of $\sim 0.5$~mrad between the $K^+$ and $K^-$ beam directions.
The remaining 65\% (8\%) of the sample correspond to $K^+$ ($K^-$) decays collected in single-beam mode. About half of the data sample was recorded with a $9.2X_0$ thick lead (Pb) bar installed between the two HOD planes and shadowing about 10\% of the LKr area. This latter setup was used for another study, as described in Ref.~\cite{la13}.

The main data set was recorded with a trigger requiring the presence of an electron~\cite{la13}, which has marginal efficiency for $K_{\pi\gamma\gamma}$ decays. Therefore a sample collected using downscaled minimum bias trigger branches requiring at least one charged particle and/or a minimum calorimetric energy deposit is used for this measurement. At least one of the following trigger conditions was required:
\begin{itemize}
\item a time coincidence of signals in the two HOD planes within the same quadrant combined with loose lower and upper limits on DCH hit multiplicity, signalling a charged particle traversing the spectrometer ($\sim 20\%$ of the data sample);
\item the above condition in coincidence with a LKr energy deposit of at least 10~GeV, signalling energy release from charged pions, electrons, or photons ($\sim 60\%$ of the data sample);
\item a signal from the NHOD detector, similarly signalling electromagnetic shower energy release ($\sim 20\%$ of the data sample).
\end{itemize}
The resulting data sample used for this measurement corresponds to about 6\% of the total beam flux.

\section{Data analysis}

\subsection{Measurement method}
\label{sec:method}

The $K_{\pi\gamma\gamma}$ decay rate is measured with respect to the normalization decay chain collected simultaneously with the same trigger logic: $K^\pm\to\pi^\pm\pi^0$ decay ($K_{2\pi}$) followed by $\pi^0\to\gamma\gamma$ decay ($\pi^0_{\gamma\gamma}$). As a consequence, the measurement does not depend on the beam flux and composition, nor the downscaling factors of the individual trigger branches and their variations throughout the data taking, provided that the time variations of the geometrical acceptances are taken into account. The similarity of the signal and normalization decay final states leads to first order cancellation of several systematic effects.

The branching ratio of $K_{\pi\gamma\gamma}$ decay can be computed as
\begin{displaymath}
{\cal B}(K_{\pi\gamma\gamma}) = \frac{N_{\pi\gamma\gamma}^\prime}{N_{2\pi}^\prime} \cdot \frac{A_{2\pi}}{A_{\pi\gamma\gamma}} \cdot \frac{\varepsilon_{2\pi}}{\varepsilon_{\pi\gamma\gamma}} \cdot {\cal B}(K_{2\pi}){\cal B}(\pi^0_{\gamma\gamma}),
\end{displaymath}
where $N_{\pi\gamma\gamma}^\prime$ and $N_{2\pi}^\prime$ are numbers of reconstructed signal and normalization events (after background subtraction), $A_{\pi\gamma\gamma}$ and $A_{2\pi}$ are the acceptances of the signal and normalization selections, and $\varepsilon_{\pi\gamma\gamma}$ and $\varepsilon_{2\pi}$ are the corresponding trigger efficiencies. The normalization mode branching ratio ${\cal B}(K_{2\pi}){\cal B}(\pi^0_{\gamma\gamma})=0.204\pm0.001$ is large and known to a good precision~\cite{pdg}.

The acceptances for the signal, normalization and background decays are evaluated with a detailed GEANT3-based~\cite{geant} Monte Carlo (MC) simulation. The signal acceptance $A_{\pi\gamma\gamma}$ is not uniform over the kinematical space, and therefore depends in general on the assumed kinematic distribution. Acceptances varied over time due to the presence of the Pb bar during part of data taking (see Section~\ref{sec:sample}), groups of LKr cells temporarily masked for hardware reasons, as well as small variations of the beam positions, directions and momenta.

Trigger efficiencies have been measured with control data samples. Due to the minimum bias trigger conditions applied and the identity of the signal and normalization
final state topologies, these efficiencies have high and similar values for the signal, normalization and background decay modes. Therefore they cancel to first order, as quantified in Section~\ref{sec:syst}.

\subsection{Event reconstruction and selection}
\label{sec:selection}

The event reconstruction is similar to that reported in Ref.~\cite{ba14}. However it is modified when needed to match the different beam and detector properties. Most of the selection criteria are common to the signal and normalization decay modes, due to their similar topologies.
\begin{itemize}
\item A $\pi^\pm$ candidate track geometrically consistent with originating from a beam $K^\pm$ decay is required. The decay vertex, reconstructed as the point of closest approach of the track and the axis of the kaon beam of the corresponding charge, should be located within a 98~m long fiducial volume contained in the vacuum tank.
\item Track impact points in the DCH, HOD and LKr calorimeter should be within their fiducial acceptances. The LKr acceptance definition includes separation by at least 6~cm from the detector edges and groups of non-instrumented or temporarily disabled cells, to reduce lateral energy leakage effects. For the data sample collected with the Pb bar installed, LKr calorimeter cells shadowed by the bar (rows 6 to 16 below the centre line) are excluded from the track geometrical acceptance, as the pions ($\pi^\pm$) traversing the bar cannot be efficiently separated from electrons ($e^\pm$) by energy deposition in the calorimeter.
\item The reconstructed track momentum is required to be between 8 and 50 GeV/$c$, which does not decrease the $K_{\pi\gamma\gamma}$ acceptance while inducing a relative loss of 5\% on the $K_{2\pi}$ acceptance. The upper momentum cut is equivalent to a lower limit on the total energy of the two photons and ensures the high efficiency of the LKr and NHOD trigger conditions; the lower momentum cut decreases the background in the $K_{\pi\gamma\gamma}$ sample.
\item The $\pi^\pm$ is identified by the ratio of energy release in the LKr calorimeter to momentum measured by the spectrometer: $E/p<0.85$. This decreases the electron contamination in the pion sample by at least a factor of 200, as measured with a sample of $K^\pm\to\pi^0 e^\pm\nu$ decays, and reduces the backgrounds from $K^\pm$ decays to electrons to a negligible level. The $\pi^\pm$ identification efficiency averaged over momentum is 98.4\%. The corresponding systematic effects are discussed in Section~\ref{sec:syst}.
\item Clusters of energy deposition in the LKr calorimeter in time with the track, separated by at least 25~cm from the track impact point and not located in the shadow of the Pb bar are considered as photon candidates. Exactly two photon candidates are required. They should be within the LKr fiducial acceptance, which is defined in the same way as for the $\pi^\pm$ candidate, except for larger separation (8~cm) from detector edges and groups of non-instrumented or temporarily disabled cells. The distance between the two candidates should be larger than 20~cm, and their energies should be above 3 GeV. These two requirements do not reduce the $K_{\pi\gamma\gamma}$ acceptance (due to the $m_{\gamma\gamma}$ cut discussed below) and lead to a relative loss of 4\% on the $K_{2\pi}$ acceptance.
\item The backgrounds result mainly from LKr cluster merging, as discussed in Section~\ref{sec:bkg}, and are characterized by larger mean lateral width of the photon candidate LKr clusters. An energy-dependent upper limit is imposed on that variable, based on measurements of width distributions of isolated electromagnetic clusters separately for data and MC simulated events. This reduces background in the $K_{\pi\gamma\gamma}$ sample by about a factor of 2, while the relative acceptance loss is below 1\% for both $K_{\pi\gamma\gamma}$ and $K_{2\pi}$ decays.
\item The reconstructed $\pi^\pm\gamma\gamma$ momentum is required to be between 70 and 78~GeV/$c$, and its component orthogonal to the axis of the kaon beam of the corresponding charge should be $p_T^2<0.5\times 10^{-3}~({\rm GeV}/c)^2$, which is consistent with the beam momentum spectrum, divergence and resolution. This leads to 1\% relative acceptance loss for both $K_{\pi\gamma\gamma}$ and $K_{2\pi}$ decays.
\item The reconstructed $\pi^\pm\gamma\gamma$ ($\pi^\pm\pi^0$) invariant mass should be between 480 and 510~MeV/$c^2$. The corresponding mass resolutions for the $K_{\pi\gamma\gamma}$ ($K_{2\pi}$) samples are 5.4 (3.3)~MeV/$c^2$.
\end{itemize}
The $K_{\pi\gamma\gamma}$ and $K_{2\pi}$ selections differ only in the di-photon invariant mass requirement.
\begin{itemize}
\item For $K_{\pi\gamma\gamma}$, the signal kinematic region is defined as $z>0.2$. Assuming a ${\cal O}(p^6)$ ChPT kinematic distribution~\cite{da96} and using the experimental input~\cite{ba14}, the expected relative $K_{\pi\gamma\gamma}$ acceptance loss is 3\%. The low $z$ region is dominated by the $\pi^0$ monochromatic line at $z=(m_{\pi^0}/m_K)^2=0.075$, which is widened by the resolution on photon energies (due to LKr energy resolution) and directions (due to LKr and spectrometer spatial resolution and beam transverse profile). As a result, the signal is not observable at low $z$, including the region below the $\pi^0$ peak. This was also the case for the previous  $K_{\pi\gamma\gamma}$~\cite{ki97,ba14} and $K_{\pi\gamma ee}$~\cite{ba08} measurements. The resolution on the $z$ variable increases from $\delta z=0.005$ at $z=0.2$ to $\delta z=0.03$ at $z_{\rm max}=0.515$.
\item For $K_{2\pi}$, the di-photon is required to be consistent with originating from a $\pi^0$ decay: $|m_{\gamma\gamma}-m_{\pi^0}|<10~{\rm MeV}/c^2$ ($0.064<z<0.086$). The mass resolution is $\delta m_{\gamma\gamma}=1.6~{\rm MeV}/c^2$ (corresponding to $\delta z=0.002$).
\end{itemize}

\begin{figure}[t]
\begin{center}
\resizebox{0.50\textwidth}{!}{\includegraphics{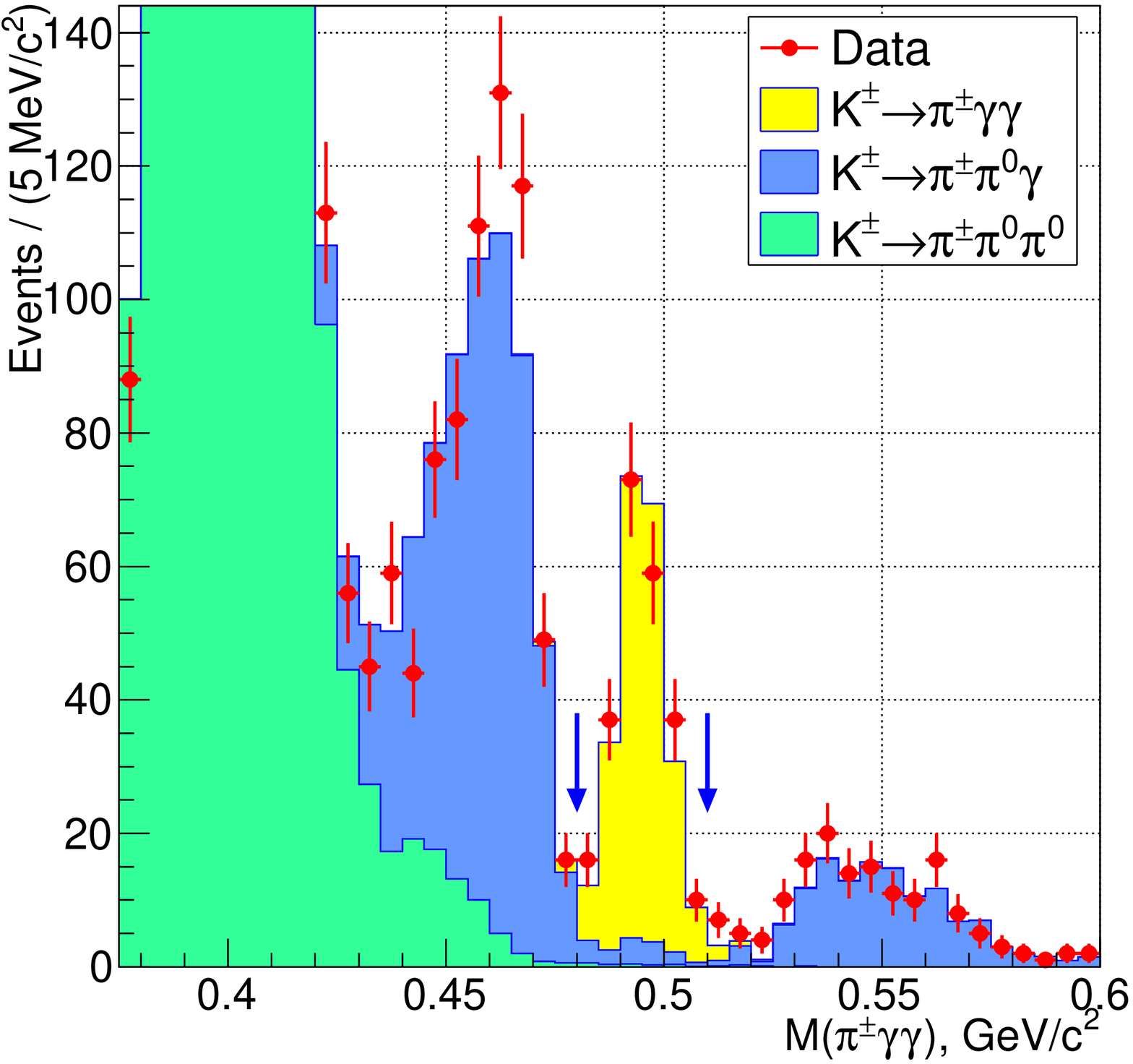}}%
\resizebox{0.50\textwidth}{!}{\includegraphics{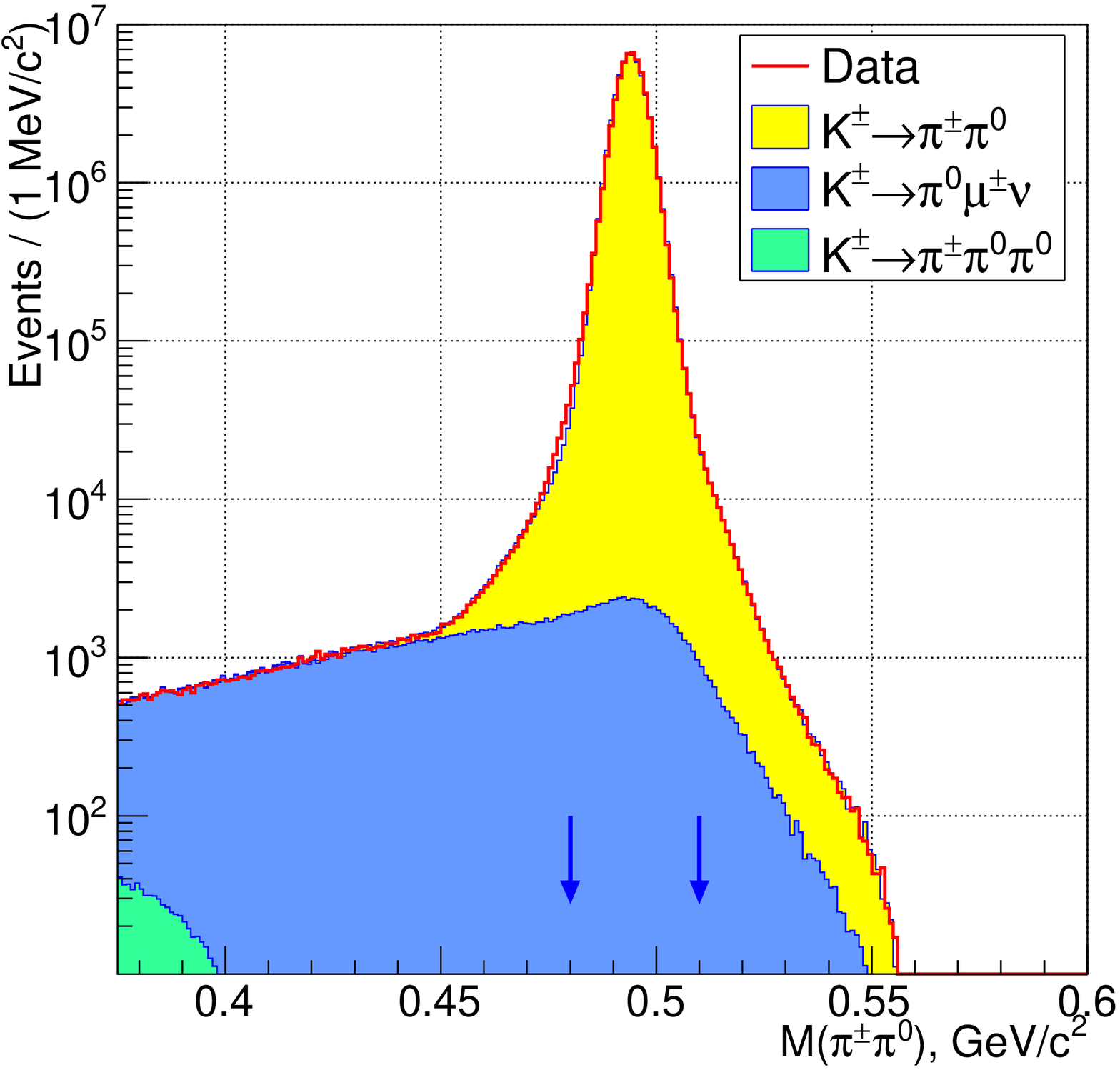}}
\put(-428,201){\bf\large (a)} \put(-201,201){\bf\large (b)}
\end{center}
\vspace{-15mm} \caption{Reconstructed invariant mass distributions of (a) $\pi^\pm\gamma\gamma$ and (b) $\pi^\pm\pi^0$ candidates compared with the sums of estimated signal and background components. The estimated $K_{\pi\gamma\gamma}$ signal corresponds to the result of a ChPT ${\cal O}(p^6)$ fit. Signal region limits are indicated with vertical arrows. The $K^\pm\to\pi^\pm\pi^0\gamma$ background contributes below, within and above the signal mass region through different mechanisms: photons missing the geometric acceptance (below), merging of photon LKr clusters (within), both combined with photon conversions in the spectrometer (above). Systematic errors on the background distributions are not indicated. The relative uncertainty on the background estimate in the $K_{\pi\gamma\gamma}$ sample is about 10\%, as discussed in Section~\ref{sec:syst}.}
\label{fig:mass}
\end{figure}

\begin{figure}[t]
\begin{center}
\resizebox{0.50\textwidth}{!}{\includegraphics{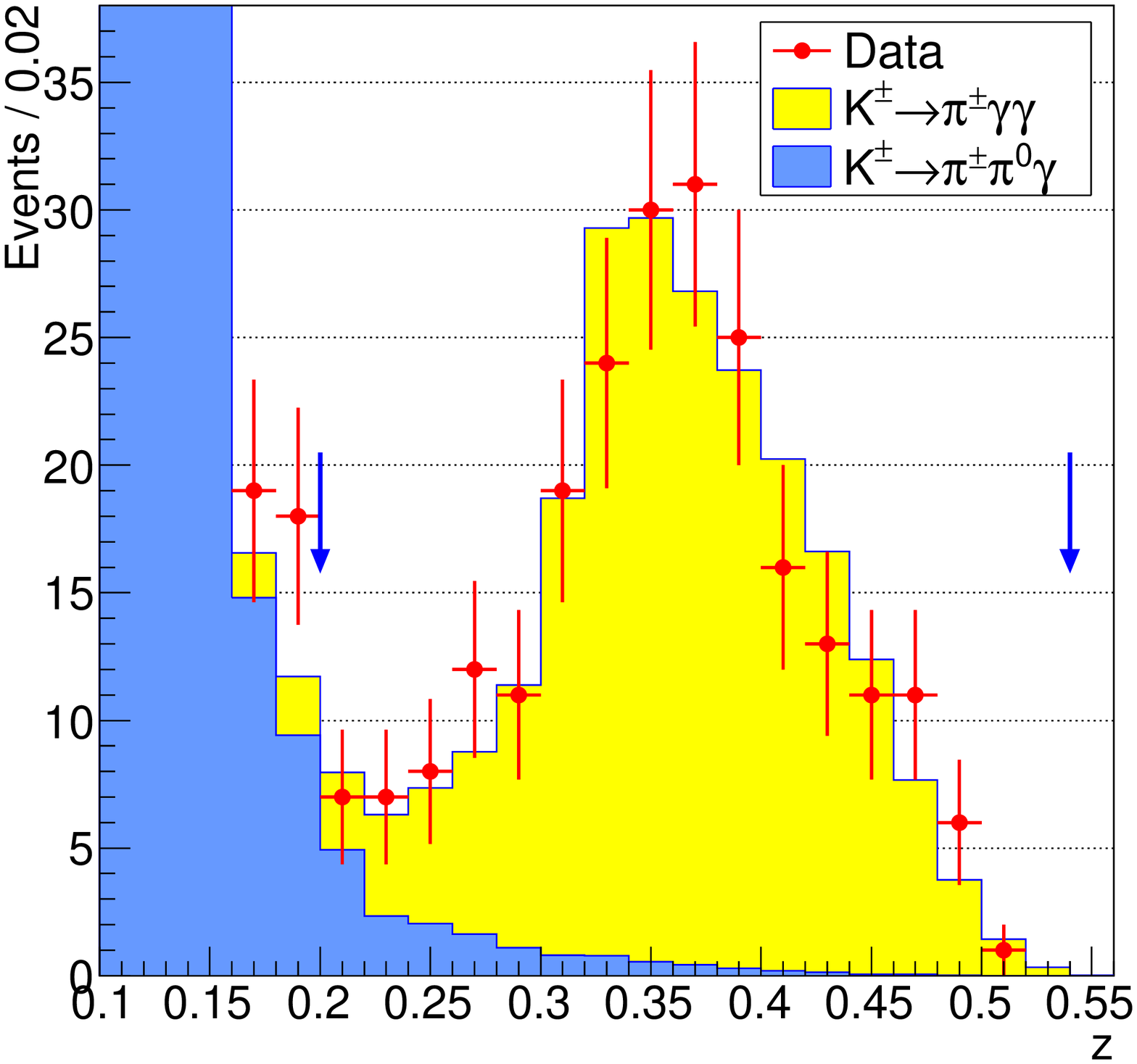}}
\end{center}
\vspace{-16mm} \caption{Reconstructed $z=(m_{\gamma\gamma}/m_K)^2$ spectrum of the $K_{\pi\gamma\gamma}$ candidates compared with the estimated contributions from the signal and the largest background $K^\pm\to\pi^\pm\pi^0\gamma$. The $K^\pm\to\pi^\pm\pi^0\pi^0$ background, which is an order of magnitude smaller, is not shown. The estimated signal corresponds to the result of a ChPT ${\cal O}(p^6)$ fit. Signal region limits are indicated with vertical arrows.}
\label{fig:z}
\end{figure}

The $\pi^\pm\gamma\gamma$ and $\pi^\pm\pi^0$ invariant mass spectra of the selected $K_{\pi\gamma\gamma}$ and $K_{2\pi}$ candidates, with the expected signal and background contributions evaluated with MC simulations, are shown in Fig.~\ref{fig:mass}. The number of $K_{\pi\gamma\gamma}$ candidates is $N_{\pi\gamma\gamma}=232$, of which 179 (53) are $K^+$ ($K^-$) decay candidates. The number of $K_{2\pi}$ candidates is $N_{2\pi} = 5.488\times 10^7$, of which $4.431~(1.057)\times 10^7$ are $K^+$ ($K^-$) decay candidates. The kaon charge composition of the sample is determined mainly by the durations of data taking periods with single $K^+$ and single $K^-$ beams. The reconstructed $z$ spectrum of the $K_{\pi\gamma\gamma}$ candidates is displayed in Fig.~\ref{fig:z}.

\subsection{Backgrounds}
\label{sec:bkg}

The only sizeable background to the normalization mode ($K_{2\pi}$, $\pi^0_{\gamma\gamma}$) comes from $K^\pm\to\pi^0\mu^\pm\nu$ decays ($K_{\mu3}$) followed by $\pi^0_{\gamma\gamma}$ decays. The background contamination is
\begin{displaymath}
R = \frac{{\cal B}(K_{\mu3})A(K_{\mu3})}{{\cal B}(K_{2\pi})A(K_{2\pi})} = 0.115\%,
\end{displaymath}
where ${\cal B}(K_{2\pi})$, ${\cal B}(K_{\mu3})$ are the nominal branching ratios of the $K^\pm$ decay modes~\cite{pdg}, and $A(K_{2\pi})=16.88\%$, $A(K_{\mu3})=0.12\%$ are the acceptances of the $K_{2\pi}$ event selection for the two decay chains evaluated with MC simulation. The quoted acceptances are obtained by averaging over the whole data sample; acceptances for the subset collected with the Pb bar (with a reduced LKr fiducial area, see Section~\ref{sec:selection}) are about two times lower than those for the subset collected without the Pb bar. The product $N_K$ of the number of $K^\pm$ decays in the fiducial decay volume in the analysed data set and the trigger efficiency for $K_{2\pi}$ sample is computed as
\begin{displaymath}
N_K = \frac{N_{2\pi}}{{\cal B}(K_{2\pi}){\cal B} (\pi^0_{\gamma\gamma})A(K_{2\pi})(1+R)}=(1.592\pm0.006)\times 10^9,
\end{displaymath}
where the uncertainty is due to the limited precision on the external input ${\cal B}(K_{2\pi})$. The number $N^B$ of background events in the $K_{\pi\gamma\gamma}$ sample is then evaluated as
\begin{displaymath}
N^B = N_K \times \sum_i {\cal B}^B_i A^B_i,
\end{displaymath}
where the sum runs over the background kaon decay modes, and ${\cal B}^B_i$ and $A^B_i$ are the corresponding branching ratios and geometrical acceptances within the $K_{\pi\gamma\gamma}$ selection. This approach relies on cancellation of the trigger efficiencies, as pointed out in Section~\ref{sec:method}.

The principal background in the $K_{\pi\gamma\gamma}$ sample comes from  $K^\pm\to\pi^\pm\pi^0\gamma$ decays followed by $\pi^0_{\gamma\gamma}$ decays. It is due to the merging of LKr energy deposition clusters produced by a photon from the $\pi^0$ decay and a photon from the parent $K^\pm$ decay, as detailed in Ref.~\cite{ba14}. This mechanism does not involve particles missing detector acceptance, therefore the relative background contamination is similar for data subsets collected with and without the Pb bar. This background is estimated with MC simulations as described in Ref.~\cite{ba14}. In particular, the dominant inner bremsstrahlung (IB) process is simulated according to Ref.~\cite{ga06}, while the smaller contributions from direct emission (DE) and interference between DE and IB are simulated using the expected ChPT phase space distributions~\cite{ch67,dafne} and the measured decay rates~\cite{ba10}. The total background from $K^\pm\to\pi^\pm\pi^0\gamma$ decays is estimated to be $15.3\pm1.1$ events, where the uncertainty is due to limited MC statistics.

Another source of background in the $K_{\pi\gamma\gamma}$ sample is due to $K^\pm\to\pi^\pm\pi^0\pi^0$ decays followed by $\pi^0_{\gamma\gamma}$ decays.
They contribute via photons missing the LKr acceptance as well as LKr cluster merging. This background is estimated to be $2.1\pm0.3$ events, where the uncertainty is also due to limited MC statistics.



\subsection{Model-independent rate measurement}
\label{sec:model-independent}

Model-independent partial $K_{\pi\gamma\gamma}$ branching ratios ${\cal B}_j$ in bins of the $z$ variable defined in Table~\ref{tab:brmi} are computed as
\begin{displaymath}
{\cal B}_j = (N_j - N^B_j) / (N_K A_j),
\end{displaymath}
where $N_j$ is the number of reconstructed $K_{\pi\gamma\gamma}$ candidates, $N^B_j$ is the estimated number of background events, $A_j$ is the signal acceptance in bin $j$, and $N_K$ is defined in Section~\ref{sec:bkg}. Trigger efficiencies nearly cancel at this stage, as discussed Section~\ref{sec:method}.

The dependence of the acceptances $A_j$ on the assumed $K_{\pi\gamma\gamma}$ kinematical distribution can be neglected with respect to the statistical uncertainties, due to the sufficiently small bin width. The $y$-dependence of the differential decay rate expected within the ChPT framework~\cite{da96,ge05} arises at next-to-leading order only, and is weak (for a fixed $z$, the relative variation of $\partial\Gamma/\partial z\partial y$ over $y$ is below 14\% for $z>0.2$ and below 6\% for $z>0.25$). The $y$-dependence of the acceptance is also weak (for a fixed $z$, the relative variation over $y$ in the range $y/y_{\rm max} < 0.9$ is below 10\%). As a result, the measurements of ${\cal B}_j$ are model-independent to a good approximation.

The values of $N_j$, $N^B_j$ and $A_j$ and the calculated ${\cal B}_j$ with their statistical uncertainties are listed in Table~\ref{tab:brmi}. The model-independent branching ratio in the kinematic region $z>0.2$ is evaluated as a sum over $z$ bins:
\begin{equation}
\label{eq:brmi}
{\cal B}_{\rm MI}(z>0.2) = \sum\limits_{j=1}^{8}{\cal B}_{j} = (1.088 \pm 0.093_{\rm stat}) \times 10^{-6}.
\end{equation}

\begin{table}[htb]
\begin{center}
\caption{Bin definition, numbers of signal and background events $N_j$ and $N^B_j$, signal acceptances $A_j$ and model-independent branching ratios ${\cal B}_j$ evaluated in $z$ bins. The quoted uncertainties are statistical. The signal acceptance reduces to zero at the endpoint $z_{\rm max}=0.515$, because the $\pi^\pm$ remains too close to the beam pipe to be detected by the drift chambers.}
\vspace{2mm}
\label{tab:brmi}
\begin{tabular}{crrcc}
\hline
Bin $z$ range & $N_j$ & $N^B_j$ & $A_j$ & ${\cal B}_j\times 10^6$\\
\hline
0.20--0.24 & 14 & 7.32 & 0.177 & $0.024 \pm 0.013$\\
0.24--0.28 & 20 & 3.83 & 0.175 & $0.058 \pm 0.016$\\
0.28--0.32 & 30 & 1.97 & 0.169 & $0.104 \pm 0.020$\\
0.32--0.36 & 54 & 1.93 & 0.160 & $0.204 \pm 0.029$\\
0.36--0.40 & 56 & 1.00 & 0.146 & $0.237 \pm 0.032$\\
0.40--0.44 & 29 & 0.57 & 0.124 & $0.144 \pm 0.027$\\
0.44--0.48 & 22 & 0.54 & 0.087 & $0.155 \pm 0.034$\\
$z>0.48$   &  7 & 0.25 & 0.026 & $0.162 \pm 0.064$\\
\hline
\end{tabular}
\end{center}
\vspace{-10mm}
\end{table}


\subsection{Measurement of ChPT parameters}
\label{sec:fits}

In the framework of the ChPT description~\cite{da96,ge05}, the $K_{\pi\gamma\gamma}$ decay receives no tree-level ${\cal O}(p^2)$ contribution. The differential decay rate at leading order ${\cal O}(p^4)$ and including ${\cal O}(p^6)$ contributions can be written as follows:
\begin{displaymath}
\frac{\partial\Gamma}{\partial y \partial z}(\hat c, y, z) = \frac{m_K}{2^9\pi^3}\left[z^2\left(|A(\hat c, z, y^2) + B(z)|^2 + |C(z)|^2 \right) + \left(y^2-\frac{1}{4}\lambda(1,r_\pi^2,z)\right)^2\left|B(z)\right|^2\right].
\end{displaymath}
Here $A(\hat c, z, y^2)$ and $B(z)$ are loop amplitudes (the latter appears at next-to-leading order and dominates the differential rate at low $z$), and $C(z)$ is a pole amplitude. The decay rate and spectrum are determined by a single a priori unknown ${\cal O}(1)$ parameter $\hat c$. The formulation of Ref.~\cite{da96} is used in this study, as it involves fewer external parameters than a similar formulation of Ref.~\cite{ge05}.

\begin{figure}[t]
\begin{center}
\resizebox{0.50\textwidth}{!}{\includegraphics{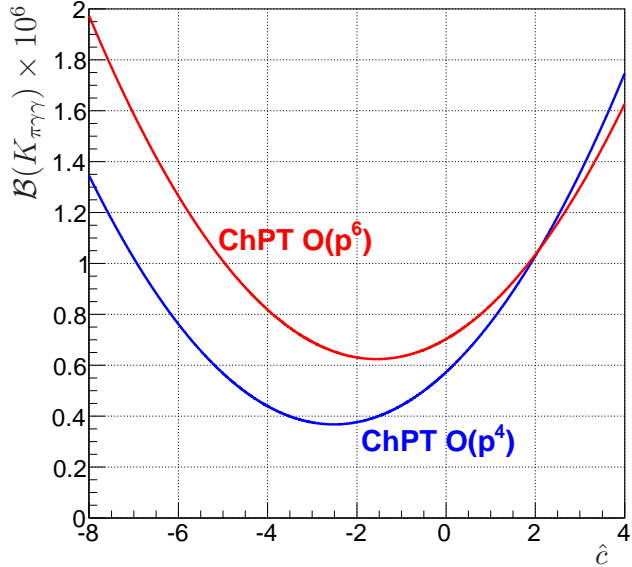}}
\put(-18,0){$\hat c$}
\put(-239,138){\rotatebox{90}{\large ${\cal B}(K_{\pi\gamma\gamma}) \times 10^6$}}
\end{center}
\vspace{-14mm} \caption{Dependence of the $K_{\pi\gamma\gamma}$ branching ratio on the $\hat c$ parameter predicted within ChPT~\cite{da96}.}
\label{fig:br_vs_c}
\end{figure}

The ChPT expectations for the differential decay rate are illustrated in Refs.~\cite{da96,ge05,ba14}: they include a cusp in the differential decay rate at the di-pion threshold $z_{\rm th}=4r_\pi^2=0.320$ generated by the pion loop amplitude. At ${\cal O}(p^6)$, a non-zero differential rate is expected at $z=0$, and this is generated by the $B(z)$ amplitude. The branching ratio in the full kinematic range is expected to be ${\cal B}(K_{\pi\gamma\gamma}) \sim 10^{-6}$, and its dependence on $\hat c$ is shown in Fig.~\ref{fig:br_vs_c}. A number of external parameters of the ChPT description are extracted from fits to experimental data~\cite{da96}. These are fixed in this study in the same way as for the NA48/2 $K_{\pi\gamma\gamma}$ measurement~\cite{ba14}.

To measure the $\hat c$ parameter in the ChPT ${\cal O}(p^4)$ and ${\cal O}(p^6)$ frameworks~\cite{da96}, fits to the reconstructed $z$ spectrum (Fig.~\ref{fig:z}) have been performed by maximizing the log-likelihood
\begin{displaymath}
\ln{\cal L} = \sum\limits_{i=1}^{17} \left[ n_i \ln m_i - m_i - \ln(n_i!)\right].
\end{displaymath}
The sum runs over bins of the reconstructed $z$ variable in the range $0.2<z<0.54$ (the bin width is $\delta z=0.02$); $n_i$ are the numbers of observed data events in the bins, and $m_i(\hat c) = m_i^{S}(\hat c) + m_i^{B}$ are the expected numbers of events for a given value of $\hat c$, including signal and background components $m_i^{S}(\hat c)$ and $m_i^{B}$. The quantities $m_i(\hat c)$ are computed using the number $N_K$ of $K^\pm$ decays in the fiducial volume measured from the normalization sample (Section~\ref{sec:bkg}), the expected $K_{\pi\gamma\gamma}$ differential decay rate for a given value of $\hat c$~\cite{da96}, and the acceptances of the $K_{\pi\gamma\gamma}$ selection for signal and backgrounds evaluated from MC simulations. The highest bin is above the $K_{\pi\gamma\gamma}$ kinematic endpoint and is populated due to resolution effects. The fits to the  ${\cal O}(p^4)$ and ${\cal O}(p^6)$ descriptions~\cite{da96} yield the following results:
\begin{displaymath}
\hat c_4 = 1.93\pm 0.26_{\rm stat}, ~~~ \hat c_6 = 2.10\pm 0.28_{\rm stat}.
\end{displaymath}
A binned Kolmogorov--Smirnov test~\cite{james} yields $p$-values of 30\% and 98\% for the ${\cal O}(p^4)$ and ${\cal O}(p^6)$ descriptions, respectively. The data are consistent with both ChPT descriptions. The $z$ spectrum corresponding to the ${\cal O}(p^6)$ fit result is shown in Fig.~\ref{fig:z}.

Within ChPT, the total decay rate has an approximately parabolic dependence on $\hat c$ (Fig.~\ref{fig:br_vs_c}), with a minimum at $\hat c=\hat c_0\approx-2$. This dependence leads to a second (local) maximum of the likelihood at the ``negative solution'' values $\hat c_4 \approx -8$, $\hat c_6 \approx -6$.  These negative solutions are however excluded on the basis of a likelihood ratio test.


\subsection{Systematic effects}
\label{sec:syst}

The largest systematic effect comes from the background estimate in the $K_{\pi\gamma\gamma}$ sample. As discussed in Section~\ref{sec:bkg}, the background is primarily due to LKr electromagnetic cluster merging. The effects of the possible differences in merging of clusters between data and MC have been studied by the variation of the cluster lateral width cut, including the removal of the cut, which approximately doubles the background. An additional stability check involving artificial merging of adjacent LKr clusters resolved by the reconstruction has been performed, as described in Ref.~\cite{ba14}. These tests have not revealed any systematic effects within their statistical sensitivity. Maximum variations of the results are conservatively considered as systematic uncertainties: $\delta {\cal B}_{\rm MI}(z>0.2)=0.027\times 10^{-6}$, $\delta\hat c_4=0.08$, $\delta\hat c_6=0.18$. The assigned uncertainties are of a statistical nature.


The signal and normalization samples have been collected with the same set of minimum bias trigger conditions. As a result, the systematic uncertainties due to trigger inefficiency are negligible as detailed below. The HOD trigger inefficiency measured with a control $K_{2\pi}$ sample collected with the NHOD trigger is $(0.4 \pm 0.1)\%$, mainly localized along the scintillator counter boundaries and with no particular pattern otherwise. Therefore it partially cancels between the $K_{\pi\gamma\gamma}$ and $K_{2\pi}$ samples, and the residual uncertainty is ${\cal O}(0.1\%)$ or below. The upper track momentum (50~GeV/$c$) and lower total momentum (70~GeV/$c$) selection conditions constrain the LKr energy deposit to be above 20 GeV, significantly above the LKr trigger threshold of 10 GeV. The corresponding LKr trigger inefficiency has been measured from a $K^\pm\to\pi^0 e^\pm\nu$ sample collected with the HOD trigger condition to be below 0.1\%. The inefficiency of the NHOD trigger has been measured as a function of photon energies from a $K_{2\pi}$ sample collected with HOD or LKr trigger conditions. The inefficiencies integrated over the data samples have been computed to be 0.15\% (0.25\%) for the $K_{\pi\gamma\gamma}$ ($K_{2\pi}$) decays. The difference is due to the higher mean photon energy and therefore higher NHOD energy deposit for $K_{\pi\gamma\gamma}$ events. A correction for this difference has been introduced for the data collected with the NHOD trigger; the residual uncertainty is negligible.

The $\pi^\pm$ identification efficiency due to the $E/p<0.85$ condition (see Section~\ref{sec:selection}) has been measured from samples of $K_{2\pi}$ and $K^\pm\to3\pi^\pm$ decays to decrease from 98.6\% at $p=8~{\rm GeV}/c$ to 98.2\% at $p=50~{\rm GeV}/c$. It is higher for the MC simulated events due to the limited precision of hadronic shower description. However it largely cancels separately for data and MC simulated samples between the signal, normalization and background channels due to its geometric uniformity and weak momentum dependence. The residual systematic bias is negligible.

The systematic uncertainties on the geometrical acceptances evaluated with MC simulations are negligible. Accidental activity effects can be neglected due to the low beam intensity. The uncertainty on the number of kaon decays in the fiducial volume due to the limited precision on the external input ${\cal B}(K_{2\pi}){\cal B}(\pi^0_{\gamma\gamma})$~\cite{pdg} leads to negligible uncertainties on the results: $\delta {\cal B}_{\rm MI}(z>0.2) = 0.004\times 10^{-6}$, $\delta\hat c_4 = \delta\hat c_6 = 0.01$.


\section{Results}

A sample of 232 $K_{\pi\gamma\gamma}$ decay candidates with an estimated background contamination of $17.4\pm1.1$ events collected with minimum bias trigger conditions by the NA62 experiment at CERN in 2007 has been analyzed. The model-independent $K^\pm_{\pi\gamma\gamma}$ branching ratio in the kinematic region $z>0.2$ is measured to be
\begin{displaymath}
{\cal B}_{\rm MI}(z>0.2) = (1.088 \pm 0.093_{\rm stat} \pm 0.027_{\rm syst}) \times 10^{-6}.
\end{displaymath}
Measurements performed separately for $K^+$ and $K^-$ decays are consistent within 1.5 standard deviations:
\begin{displaymath}
{\cal B}^+_{\rm MI}(z>0.2) = (1.010 \pm 0.098_{\rm stat}) \times 10^{-6},~~~
{\cal B}^-_{\rm MI}(z>0.2) = (1.417 \pm 0.256_{\rm stat}) \times 10^{-6}.
\end{displaymath}
The observed decay spectrum agrees with the ChPT expectations. The values of the $\hat{c}$ parameter in the framework of the ChPT ${\cal O}(p^4)$ and ${\cal O}(p^6)$ parameterizations~\cite{da96} have been obtained from fits to the data $z$ spectrum:
\begin{eqnarray}
\hat c_4 &=& 1.93 \pm 0.26_{\rm stat} \pm 0.08_{\rm syst},\nonumber\\
\hat c_6 &=& 2.10 \pm 0.28_{\rm stat} \pm 0.18_{\rm syst}.\nonumber
\end{eqnarray}
The data are insufficient to discriminate between the ${\cal O}(p^4)$ and ${\cal O}(p^6)$ parameterizations. The measured value of $\hat c_6$ translates into the following model-dependent branching ratio in the full kinematic range, obtained by integration of the ChPT ${\cal O}(p^6)$ differential decay rate:
\begin{displaymath}
{\cal B}_{\rm ChPT} = (1.058 \pm 0.066_{\rm stat} \pm 0.044_{\rm syst}) \times 10^{-6}.
\end{displaymath}
The statistical error of ${\cal B}_{\rm ChPT}$ is smaller than that of ${\cal B}_{\rm MI}$ because the low acceptance in the high $z$ bins leads to large statistical errors of ${\cal B}_j$ (see Table~\ref{tab:brmi}), which propagate directly into ${\cal B}_{\rm MI}$ according to Eq.~(\ref{eq:brmi}), while having a small influence on the fitting procedure used to obtain ${\cal B}_{\rm ChPT}$.


\section{Discussion}

\boldmath
\subsection{Combination with the NA48/2 $K_{\pi\gamma\gamma}$ results}
\unboldmath
\label{sec:combination}

A combination of the present results with those from the NA48/2 $K^\pm_{\pi\gamma\gamma}$ measurement~\cite{ba14} has been performed. Systematic uncertainties on the combined results are dominated by those due to background subtraction. They have been estimated by studying the stability of the combined results
with respect to variation of the selection conditions applied separately
to the independent NA48/2 and NA62 data samples.

\begin{figure}[t]
\begin{center}
\resizebox{0.50\textwidth}{!}{\includegraphics{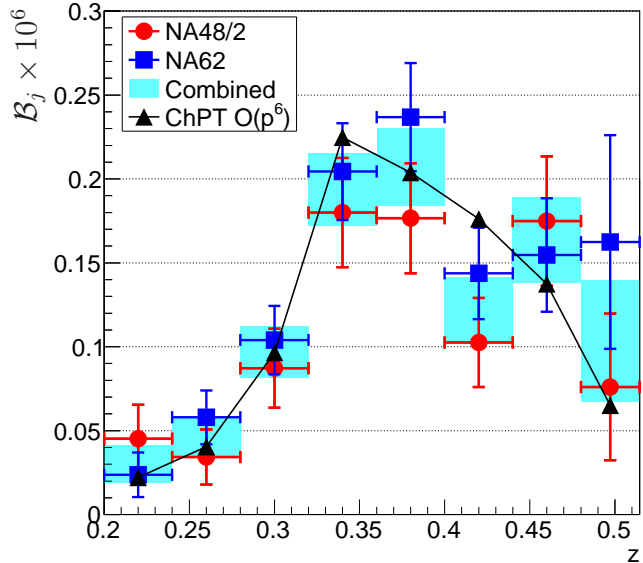}}
\put(-240,170){\rotatebox{90}{\large ${\cal B}_j \times 10^6$}}
\end{center}
\vspace{-16mm} \caption{Measurements of the model-independent $K_{\pi\gamma\gamma}$ branching ratios ${\cal B}_j$ in $z$ bins: NA48/2~\cite{ba14}, present result and the combination of the two. The horizontal bars indicate the bin widths, while the vertical bars indicate the statistical errors. The values of ${\cal B}_j$ computed within ChPT ${\cal O}(p^6)$ formulation~\cite{da96}, obtained by integration of the ChPT differential decay rate for the central value of the combined measurement $\hat c_6=1.86$ over the bin width, are also shown. The lines connecting the markers are drawn to guide the eye.}
\label{fig:brmi}
\end{figure}

The combined measurements of model-independent branching ratios ${\cal B}_j$ in $z$ bins with their statistical uncertainties are presented in Table~\ref{tab:brmi-combi}. The NA48/2 and NA62 measurements of ${\cal B}_j$ are in agreement, as seen in Fig.~\ref{fig:brmi}. The model-independent branching ratio ${\cal B}_{\rm MI}(z>0.2)$ obtained by summing the combined values of ${\cal B}_j$ is
\begin{displaymath}
{\cal B}_{\rm MI}(z>0.2) = (0.965 \pm 0.061_{\rm stat} \pm 0.014_{\rm syst}) \times 10^{-6}.
\end{displaymath}

\begin{table}[htb]
\begin{center}
\vspace{-3mm}
\caption{Combined results of this analysis and NA48/2 values~\cite{ba14} for the model-independent $K_{\pi\gamma\gamma}$ branching ratio ${\cal B}_j$ in $z$ bins. The quoted errors are statistical only.}
\vspace{2mm}
\label{tab:brmi-combi}
\begin{tabular}{cc}
\hline
Bin $z$ range & ${\cal B}_j\times 10^6$\\
\hline
0.20--0.24 & $0.030 \pm 0.011$\\
0.24--0.28 & $0.046 \pm 0.011$\\
0.28--0.32 & $0.097 \pm 0.015$\\
0.32--0.36 & $0.194 \pm 0.022$\\
0.36--0.40 & $0.207 \pm 0.023$\\
0.40--0.44 & $0.123 \pm 0.019$\\
0.44--0.48 & $0.164 \pm 0.025$\\
$z>0.48$   & $0.104 \pm 0.036$\\
\hline
\end{tabular}
\end{center}
\vspace{-10mm}
\end{table}

The combined ${\cal B}_{\rm MI}^\pm(z>0.2)$ measurements separately for $K^+$ and $K^-$ decays, obtained by averaging the NA48/2 and NA62 ${\cal B}_{\rm MI}^{\pm}(z>0.2)$ results, are
\begin{displaymath}
{\cal B}^+_{\rm MI}(z>0.2) = (0.951 \pm 0.072_{\rm stat}) \times 10^{-6},~~~
{\cal B}^-_{\rm MI}(z>0.2) = (1.004 \pm 0.127_{\rm stat}) \times 10^{-6}.
\end{displaymath}
These values are consistent: the charge asymmetry of the decay rate is $\Delta(K_{\pi\gamma\gamma})=({\cal B}^+_{\rm MI}-{\cal B}^-_{\rm MI})/({\cal B}^+_{\rm MI}+{\cal B}^-_{\rm MI})=-0.03\pm0.07$, where the sub-dominant systematic uncertainties are neglected.

The NA48/2 and NA62 measurements of the ChPT parameter $\hat c$ consider the same set of external parameters. The combination of the two results is therefore straightforward, and the results are
\begin{displaymath}
\hat c_4 = 1.72 \pm 0.20_{\rm stat} \pm 0.06_{\rm syst}, ~~ \hat c_6 = 1.86 \pm 0.23_{\rm stat} \pm 0.11_{\rm syst},
\end{displaymath}
where the second error is the experimental systematic uncertainty. Integration of the ${\cal O}(p^6)$ differential decay rate for the above value of $\hat c_6$ over the whole physical region of the kinematic variables leads to the following branching ratio:
\begin{displaymath}
{\cal B}_{\rm ChPT} = (1.003 \pm 0.051_{\rm stat} \pm 0.024_{\rm syst}) \times 10^{-6} = (1.003\pm0.056) \times 10^{-6}.
\end{displaymath}
The corresponding model-dependent values of ${\cal B}_{\rm ChPT}$ in $z$ bins are displayed in Fig.~\ref{fig:brmi}.

\subsection{Comparison with earlier measurements of ChPT parameters}

The measurements of the $\hat c$ parameter in the ChPT framework~\cite{da96,ga99} published before this analysis and Ref.~\cite{ba14} are:
\begin{itemize}
\item $\hat c_4$ and $\hat c_6$ measurements from a sample of 31 $K^+_{\pi\gamma\gamma}$ decay candidates recorded by the BNL E787 experiment~\cite{ki97};
\item $\hat c_6$ measurement from a sample 120 $K_{\pi\gamma ee}^\pm$ decay candidates recorded by the NA48/2 experiment~\cite{ba08}.
\end{itemize}
The values of external parameters considered for these measurements differ from those used to obtain the results reported in Section~\ref{sec:combination}, as summarized in Table~\ref{tab:extpar}.\footnote{The exact values of the external parameters considered for the earlier measurements are not explicitly given in Refs.~\cite{ki97, ba08}, but were provided by the corresponding authors.}


\begin{table}[htb]
\begin{center}
\caption{Values of the external parameters considered for $\hat c$ measurements from $K_{\pi\gamma\gamma}$ and $K_{\pi\gamma ee}$ decays. The notation is introduced in Refs.~\cite{da96,ci12,bi03}.}
\vspace{2mm}
\label{tab:extpar}
\begin{tabular}{crrr}
\hline
Measurement & BNL E787~\cite{ki97} & NA48/2~\cite{ba08} & NA48/2~\cite{ba14} and\\
&&&present analysis\\
\hline
Decay mode & $K_{\pi\gamma\gamma}^+$ & $K_{\pi\gamma ee}^\pm$ &
$K_{\pi\gamma\gamma}^\pm$\\
\hline
$G_8m_K^2\times 10^6$ &     2.24 &   2.210 & 2.202   \\
$\alpha_1\times 10^8$ &  $91.71$ &  $91.7$ & $93.16$ \\
$\alpha_3\times 10^8$ &  $-7.36$ &  $-7.4$ & $-6.72$ \\
$\beta_1\times 10^8$  & $-25.68$ & $-25.7$ & $-27.06$\\
$\beta_3\times 10^8$  &  $-2.43$ &  $-2.4$ & $-2.22$ \\
$\gamma_3\times 10^8$ &   $2.26$ &   $2.3$ &  $2.95$  \\
$\zeta_1\times 10^8$  &  $-0.47$ &  $-0.5$ & $-0.40$ \\
$\xi_1\times 10^8$    &  $-1.51$ &  $-1.5$ & $-1.83$ \\
$\eta_i~(i=1;2;3)$    &      $0$ &     $0$ &  $0$ \\
\hline
\end{tabular}
\end{center}
\vspace{-10mm}
\end{table}

The $G_8$ parameter enters both ${\cal O}(p^4)$ and ${\cal O}(p^6)$ descriptions. It is fixed according to Ref.~\cite{ci12} in this study. Considering higher values of $G_8$ used for the earlier E787 (NA48/2) measurements modifies the results reported in Section~\ref{sec:combination} by $\Delta\hat c_4=-0.12(-0.02)$, $\Delta\hat c_6=-0.04(-0.01)$, respectively. The shift is negative because the ChPT decay rate for the experimentally established ``positive solution'' ($\hat c>\hat c_0\approx -2$) is characterised by \mbox{$\partial{\cal B}/\partial\hat c>0$} and $\partial{\cal B}/\partial G_8>0$.

The $K_{3\pi}$ amplitude parameters enter the ${\cal O}(p^6)$ formulation. They are fixed according to Ref.~\cite{bi03} in this study. Both earlier measurements employed a set of parameters obtained from another fit~\cite{ka91}. The corresponding shift of the result quoted in Section~\ref{sec:combination} is $\Delta\hat c_6=-0.26$, primarily due to the sensitivity to the $\xi_1$ parameter.

This study, similarly to the previous measurements, considers zero values of the polynomial contributions entering the ${\cal O}(p^6)$ formulation ($\eta_i=0$). The parameter $\hat c$ enters the differential decay rate via a linear combination
\begin{equation}
\label{eq:cprime}
\hat c^* = \hat c - 2(m_\pi/m_K)^2\eta_1 - 2\eta_2 - 2\eta_3.
\end{equation}
Assuming $\eta_i=0$ is equivalent to measuring $\hat c^*$. The value of $\hat c$ can be computed for any set of $\eta_i$ from the measured $\hat c^*$ using the above relation.

The combined $\hat c$ measurements from this analysis and Ref.~\cite{ba14} obtained using the set of external parameters considered in Ref.~\cite{ki97} are
\begin{eqnarray}
\hat c_4^\prime &=& 1.60\pm0.20_{\rm stat}\pm0.06_{\rm syst},\nonumber\\
\hat c_6^\prime &=& 1.56\pm0.23_{\rm stat}\pm0.11_{\rm syst}.\nonumber
\end{eqnarray}
They agree with the E787 results $\hat c_4=1.6\pm 0.6$, $\hat c_6=1.8\pm0.6$~\cite{ki97}. Similarly, the combined $\hat c_6$ measurement obtained using the set of external parameters considered in Ref.~\cite{ba08} is
\begin{displaymath}
\hat c_6^{\prime\prime} = 1.59\pm0.23_{\rm stat}\pm0.11_{\rm syst},
\end{displaymath}
which agrees to 1.3 standard deviations with the NA48/2 measurement $\hat c_6 = 0.90\pm0.45$ from a $K^\pm_{\pi\gamma ee}$ sample~\cite{ba08}. However, a comparison of $\hat c$ measurements obtained from different decay modes ($K_{\pi\gamma\gamma}$ and $K_{\pi\gamma ee}$) might be affected by additional external uncertainties.

The branching ratio in the full kinematic range ${\cal B}_{\rm ChPT}$ obtained assuming the ChPT ${\cal O}(p^6)$ description, reported in Section~\ref{sec:combination}, has a negligible sensitivity to the above differences of external parameters. It agrees with the E787 result ${\cal B}_{\rm ChPT}=(1.1\pm0.3\pm0.1)\times 10^{-6}$~\cite{ki97}, and is 5 times more precise. It is also in agreement with an early prediction for the total decay rate $\Gamma(K_{\pi\gamma\gamma})=76~{\rm s}^{-1}$~\cite{se72} which, considering a mean $K^\pm$ lifetime of $\tau_K=(1.2380\pm0.0021)\times 10^{-8}$~s~\cite{pdg}, translates into ${\cal B}(K_{\pi\gamma\gamma})=\tau_K\Gamma(K_{\pi\gamma\gamma}) = (0.941\pm0.002)\times 10^{-6}$.


\subsection{External uncertainties on ChPT parameters}

The expressions for the ${\cal O}(p^4)$ loop amplitude $A(\hat c, z)$ differ between the ChPT formulations of Ref.~\cite{da96} and Ref.~\cite{ge05}: the latter includes non-octet $G_{27}$ terms. The difference between the fit results derived from the two formulations is $\delta\hat c_4=0.25$. The expression for the ${\cal O}(p^6)$ loop amplitude with non-octet terms is not available in the literature.

Another difference between the two ChPT formulations lies in the computation of the pole amplitude $C(z)$. Its contribution to the total decay rate is sub-dominant (${\cal B}_{\rm pole}=0.05\times 10^{-6}$ according to~\cite{da96}; ${\cal B}_{\rm pole}<0.03\times 10^{-6}$ according to~\cite{ge05}), and the presently available data sample is not large enough to distinguish between the two formulations. The difference between the fit results with and without inclusion of the $C(z)$ amplitude gives conservative estimates of the corresponding external uncertainties: $\delta\hat c_4=0.16$, $\delta\hat c_6=0.18$.

The uncertainties on the $K_{3\pi}$ decay amplitude parameters induce an error on the measured $\hat c_6$ parameter. By far, the largest contribution comes from $\xi_1$: an uncertainty of $\delta\xi_1=0.30\times 10^{-8}$~\cite{bi03} translates into $\delta \hat c_6 = 0.30$. The uncertainty on $\xi_1$ could be reduced by considering the precision measurements of the $K_{3\pi}$ decay amplitudes~\cite{ba07_k3pi, ba09_k3pi}. The errors on the measured $\hat c_6$ due to the assumption $\eta_i=0$ can be evaluated from Eq.~(\ref{eq:cprime}). The uncertainties on the external parameters $\eta_i$ are not available in the literature.

The total external error is comparable to or larger than the achieved experimental precision.

\section*{Summary}

A model-independent measurement of the $K_{\pi\gamma\gamma}$ decay rate and fits to the ChPT description have been performed. The results have been combined with those from a recent $K_{\pi\gamma\gamma}$ measurement by the NA48/2 collaboration. The results of the combination are as follows. The model-independent branching ratio in a limited kinematic range is ${\cal B}_{\rm MI}(z>0.2) = (0.965 \pm 0.063) \times 10^{-6}$ and its charge asymmetry is $\Delta(K_{\pi\gamma\gamma})=-0.03\pm0.07$. The observed decay spectrum agrees with the ChPT description, and ChPT parameters measured within the considered formulations are $\hat c_4 = 1.72 \pm 0.21$ and $\hat c_6 = 1.86 \pm 0.25$. The branching ratio in the full kinematic range assuming the ${\cal O}(p^6)$ description is ${\cal B}_{\rm ChPT} = (1.003\pm0.056) \times 10^{-6}$. The uncertainties are dominated by the statistical errors while including a small experimental systematic contribution.

These results agree with earlier data and improve significantly on previous experimental knowledge. The obtained experimental precision ($\delta \hat c\approx 0.2$, $\delta{\cal B}_{\rm ChPT}/{\cal B}_{\rm ChPT}\approx 5\%$) may prompt more refined theoretical studies to constrain better the external parameters whose uncertainties are now larger than the experimental errors.


\section*{Acknowledgements}

It is a pleasure to express our appreciation to the staff of the CERN laboratory and the technical staff of the participating laboratories and universities for their efforts in the operation of the experiment and data processing. We are grateful to Giancarlo d'Ambrosio, John Fry and Jorge Portol\'es for fruitful discussions.



\end{document}